\documentstyle[prb,aps]{revtex}
\tightenlines

\begin{document}
\author{M.A.Teplov, Yu.A.Sakhratov, A.V.Dooglav, A.V.Egorov,}
\address{E.V.Krjukov, O.P.Zaitsev\\
Kazan State University, 420008 Kazan, Russia}
\title{Stripe motion in CuO$_2$ planes of Y$_{1-x}$Pr$_x$Ba$_2$Cu$_3$O$_7$ as seen
from the Cu(2) NQR}
\maketitle

\begin{abstract}
Pulsed NQR at the frequencies of 28-33 MHz has been used to study copper NQR
spectra in YBa$_2$Cu$_3$O$_7$, TmBa$_2$Cu$_3$O$_7$ and Y$_{0.9}$Pr$_{0.1}$Ba$%
_2$Cu$_3$O$_7$ compounds at temperatures of 4.2-200K. Quantitative analysis
of the spectra has shown that the ''plane'' Cu(2) spectra shape is well
described by using a model of 1D correlations of charge and spin
distribution in CuO$_2$ planes (stripe correlations). In the undoped
superconductors the charge-spin stripe structure moves fast in the CuO$_2$
planes, but doping the YBa$_2$Cu$_3$O$_7$ lattice with praseodymium slows
this motion down.
\end{abstract}

The hypothesis of 1D ordering of charges and spins in a particular
configuration of stripes in CuO$_2$ planes of RBa$_2$Cu$_3$O$_7$
superconductors has been suggested two years ago [1] and was used later to
explain NMR and NQR data for TmBa$_2$Cu$_3$O$_{6+x}$ and TmBa$_2$Cu$_4$O$_8$
[2], but up to now it has not been corroborated directly in the shape of
Cu(2) NQR spectra. In this paper we present such corroboration based on the
analysis of the Cu(2) NQR spectra for Y$_{1-x}$Pr$_x$Ba$_2$Cu$_3$O$_7$ (x=0,
0.1) compounds. The YBa$_2$Cu$_3$O$_7$ (YBCO) and Y$_{0.9}$Pr$_{0.1}$Ba$_2$Cu%
$_3$O$_7$ (YPBCO) samples prepared by solid-state reaction method [3] were
kindly placed at our disposal by Dr. Y.Xu and Dr. H.Luetgemeier
(Forschungszentrum Juelich, Germany). The critical temperatures found from
susceptibility measurements at the frequency of 1kHz appeared to be T$_c$%
(onset)=92.5 and 86 K, respectively. For comparison, besides YBCO and YPBCO,
the overdoped TmBa$_2$Cu$_3$O$_7$ (TmBCO) compound with the critical
temperature of 91.5 K has been studied [2]. Home-built spin-echo coherent
pulsed spectrometer was used for copper NQR spectra measurements.

The examples of YBCO and YBPCO spectra are shown in Fig.1. It is seen
(Fig.1a,c) that both spectra, except the relatively narrow $^{63}$Cu(2) and $%
^{65}$Cu(2) NQR lines, have a broad ''pedestal''. Representing these spectra
in a logarythmic scale (Fig.1b,d) one can clearly see the asymmetry of the
narrow lines, i.e., complex composition of these lines also, in the sample
doped with praseodymium. We managed to get the best fit of the YBCO spectra
to six Gaussian curves, three for each isotope: the narrow Gaussian line $%
A^{\prime }$, the broad one $A^{\prime \prime }$ and the pedestal $P$. The
frequencies $\nu _i$ and mean square linewidths $\Delta \nu _i$ of each pair
of isotope lines were supposed to be related by the ratio of nuclear
quadrupole moments $\alpha $=$^{65}$Q/$^{63}$Q ($^{65}\nu _i=\alpha ^{63}\nu
_i$, $^{65}\Delta \nu _i=\alpha ^{63}\Delta \nu _i$, $i=A^{\prime
},A^{\prime \prime },P$). Thus the fitting function involved 10 parameters:
three line frequencies $^{63}\nu _i$, three linewidths $^{63}\Delta \nu _i$,
three line intensities $^{63}a_i$ and the ratio $b=^{65}a_i/^{63}a_i$. While
fitting the YBCO spectra taken at temperatures 200-4.2K it was found that
the pedestal linewidth ($1.5\pm 0.1$ MHz) did not depend on temperature, so
its value was taken fixed at 1.5 MHz during the final fitting procedure (the
results of which is shown in Fig.1a and Fig.2a,b,c) thus reducing the total
number of fitting parameters to 9. It should be mentioned here that the
presence of pedestal is typical for copper NQR spectra in 1-2-3-7 compounds
(but not in 1-2-4-8 ones [2]). We observed it in the spectra of Tm1-2-3-7
and Y1-2-3-7 samples prepared in different laboratories [4]; in some of them
the pedestal looked like several poorly resolved overlapping lines and, in
particular, contained the broad NQR line of two-fold coordinated ''chain'' $%
^{63}$Cu(1)$_2$ copper at the frequency of 30.1MHz. We suppose the spectrum $%
P$ to arise from the Cu(1) and Cu(2) centers located in the areas with
partly disordered oxygen sublattice of CuO planes and, respectively, with
reduced (and locally inhomogeneous) hole concentration in CuO$_2$ planes.
Extracting the sum ($A^{\prime }+A^{\prime \prime }$) from the observed
spectrum we obtain the spectrum of ''good'' superconductor, free (or almost
free) of crystal structure defects. Actually this procedure allowed us to
obtain (for the first time, to our knowledge) a quantitative description of
Cu(2) NQR line shape for the superconductor with the 1-2-3-7 orthorhombic
structure. A comparative analysis of spectra for YBCO (Fig.1a and 2a,b,c)
and TmBCO [4] has shown that the lineshape is intermediate between Gaussian
and Lorentzian and for both samples the same relation holds between the
parameters describing the lines $A^{\prime }$ and $A^{\prime \prime }$. In
particular, in the temperature range of 100-150K we have obtained $\nu
_{A^{\prime }}=\nu _{A^{\prime \prime }}$, $\Delta \nu _{A^{\prime \prime
}}/\Delta \nu _{A^{\prime }}=2.4(1)$ and $a_{A^{\prime \prime
}}/a_{A^{\prime }}=1.64(5)$, although the linewidths $\Delta \nu _i$ for
TmBCO has appeared to be 25-30\% larger than those for YBCO.

In order to describe the copper NQR spectra shape for YPBCO sample we used
the fitting function containing four Gaussians for each isotope: besides the
pedestal (its linewidth $^{63}\Delta \nu _P$ was again accepted to be equal
to 1.5MHz), the central line $A$ (since the frequencies $\nu _{A^{\prime }}$
and $\nu _{A^{\prime \prime }}$ are close in YBCO, we considered it possible
to restrict the description of this line in YPBCO to one line) and the
satellite $C$ at the right slope of the spectrum, we have introduced the
line $B$ located at the frequency $\nu _B<\nu _A$ in the spectrum (the
presence of this line is revealed by the bulging left slope of the $^{63}$Cu
line in Fig.1d). Thus the fitting function contained 12 parameters: four
frequencies $^{63}\nu _i$, three linewidths $^{63}\Delta \nu _i$, four
intensities $^{63}a_i$ and the intensity ratio $b$=$^{65}a_i$/$^{63}a_i$.
The results of fitting of YBCO spectra taken at 200-4.2K are given in
Fig.2d,e,f. Let us point out and discuss the main peculiarities of copper
NQR spectra for YBCO, TmBCO and YPBCO samples.

1. The integral intensity of the pedestal $P$ is the same for YBCO and TmBCO
and equals approximately to 1/3, but for YPBCO it increases up to $\sim 1/2$%
; this confirms our assumption that the component $P$ of the spectra belongs
to partly disordered phase of the compounds studied.

Other remarks concern with ''pure'' spectra of Cu(2) NQR characterised by
the components $A^{\prime }$, $A^{\prime \prime }$ (YBCO, TmBCO) and $A$, $B$%
, $C$ (YPBCO).

2. In the spectra of YBCO (Fig.2a,b,c) and TmBCO [4] at temperatures
100-200K the frequencies $\nu _{A^{\prime }}$ and $\nu _{A^{\prime \prime }}$
are approximately equal to each other, and $\nu _{A^{\prime }}<\nu
_{A^{\prime \prime }}$ at $T<T_c$ . In general the difference of these
frequencies is small ($<0.2$\%), so that for rough estimates it is possible
to assume that the sum ($A^{\prime }+A^{\prime \prime }$) describes a single
Cu(2) NQR line in YBCO and TmBCO samples.

3. The temperature dependence of the frequency of this line in the YBCO
spectrum coincides with the temperature dependence of the line $A$ in the
spectrum of YPBCO sample (cf. Figs.2a and 2d). This allows us to ascribe the
line $A$ (Fig.2d) to the Cu(2) nuclei located far away from Pr atoms and not
influenced by them. Let us call these nuclei as ''remote''.

4. The main result of this paper is represented by the temperature
dependences of the parameters of the lines $A,B,C$ for YPBCO samples
(Fig.2d,e,f). We ascribe the lines $B$ and $C$ which are absent in the YBCO
spectrum (Fig.2a) to the ''neighboring'' nuclei, i.e., to those located not
far from Pr atoms. Comparing the integral intensities $a_i$ of the lines at
temperatures above T$_c$ (when the spectra are free of possible distortions
due to different penetration depth of the RF field $H_1$ into areas with
different local concentration of holes), we obtain the ratio of the mean
intensities $<a_B>:<a_C>$=2:1, just the same as that following from the
model of quasi-1D ordering of charges and spins in the CuO$_2$ planes
(stripe model) [1]. Furthermore, we have noticed that the lines $B$ and $C$
are located asymmetrically with respect to the line $A$, but in such a way
that $(\nu _Ba_B+\nu _Ca_C)/(a_B+a_C)=\nu _A$ ( open squares in Fig.2d).
These two facts immediately suggest the idea that both spectra, that for
''remote'' nuclei ($A$) and ''neighboring'' ones ($B$+$C$), belong to the
stripes moving in the CuO$_2$ planes, the only difference between them being
the different rate of motion: for ''remote'' nuclei the case of fast motion
is realised while for ''neighboring'' - that of slow motion. Two types of
Cu(2) centers are distinguished in the model [1] - those located in the
center of the stripe (type $C$, the hole density on the oxygen ligands is
high) and at the stripe boundaries (type $B$, the hole density is low), so
that at the optimal doping of the CuO$_2$ planes by holes and at the dense
packing of the stripes the amount of centers $B$ is twice as large as that
of centers $C$. The shape of Cu(2) NQR spectrum in the system of moving
stripes can be described by the following function [5]: 
\begin{equation}
\label{eq1}I(\omega ,\Omega )=\frac{W_BW_C(\omega _B-\omega _C)^2(\tau
_B+\tau _C)\tau _B\tau _C}{\left[ \tau _B\tau _C(\omega -\omega _B)(\omega
-\omega _C)\right] ^2+\left[ \tau _B(\omega -\omega _B)+\tau _C(\omega
-\omega _C)\right] ^2}, 
\end{equation}
where $W_i$ is the probability to find the nucleus in the $i$-th state, $%
\tau _i$ and $\omega _i=\Omega +\Delta _i$ are the lifetime of this state
and the corresponding NQR frequency, respectively. Actually the frequency $%
\Omega $ is randomly distributed near the mean value $<\Omega >=\omega _0$
(the quadrupole broadening due to the lattice imperfections), thus the
spectrum shape is obtained by averaging Eq.(\ref{eq1}) with the Gaussian
distribution of $\Omega :$%
\begin{equation}
\label{eq2}S(\omega )\sim \int I(\omega ,\Omega )\exp \left[ -(\Omega
-\omega _0)^2/2\sigma ^2\right] d\Omega \text{ .} 
\end{equation}

The $^{63}$Cu(2) NQR spectrum taken at 100K is shown by circles in Fig.3a,
the squares in Fig.3b depict the spectrum $^{63}$($B$+$C$) for $^{63}$Cu
isotope obtained by subtraction of three lines ($^{63}A$, $^{63}P$ and the $%
P $ component for $^{65}$Cu isotope) from the experimental spectrum
(Fig.3a), the triangles in Fig.3c display the $^{63}$($A$+$B$+$C$) spectrum.
The solid line in Fig.3b represents the calculated spectrum $S_{B+C}(\omega
) $ obtained at the following values of parameters: $\Delta _B=-\Delta
=-2\pi \cdot 217\cdot 10^3$ s$^{-1}$, $\Delta _C=2\Delta $, $\tau _C=\tau
=2.9\cdot 10^{-6}$ s, $\tau _B=2\tau $, $\omega _0=2\pi \cdot 31.465\cdot
10^6$ s$^{-1} $, $\sigma =2\pi \cdot 159\cdot 10^3$ s$^{-1}$. One can see
that the experimental spectrum $S_{B+C}$ of ''neighboring'' nuclei and total
NQR spectrum ($S_A+S_{B+C}$) are described very well in the frame of our
model of moving stripes (Fig.3b,c). The spectrum of ''remote'' nuclei
(dashed line $S_A$ in Fig.3c) is obtained at the same $\Delta _B$, $\Delta _C
$ values but when the short lifetime $\tau $ ($<10^{-7}$ s) is assumed; the
frequency of the line $A$ appears to be 0.09\% higher (this corresponds to
higher mean concentration of the holes), and the quadrupolar linewidth -
25\% less than these parameters for ''neighboring'' nuclei. Thus far away
from impurity Pr ions the motion rate of the stripes is high, the
corresponding lifetime $\tau $ is very small and cannot be evaluated in the
present experiment.

Two conclusions can be inferred from above: 1) the fast motion of stripes in
the CuO$_2$ planes seems to be necessary for cuprates to superconduct; 2) Pr
doping leads to the pinning of the stripes resulting in a suppression of
superconductivity in Y$_{1-x}$Pr$_x$Ba$_2$Cu$_3$O$_7$. The 1D correlations
in charge and spin distribution remain valid in the latter case too, but
they become static in character.

We are indebted to Dr. Y.Xu and Dr. H.Luetgemeier for the samples placed at
our disposal. Valuable discussions with V.A.Atsarkin, M.V.Eremin and
V.F.Gantmakher are gratefully aknowledged. This work was conducted as a part
of the Russian Scientific and Technical Program ''Current Problems in the
Physics of Condensed Matter'' (sub-program ''Superconductivity'', Project
No.94029) and the International Russian-German Program ''Spectroscopy of
High-T$_c$ Superconductors''.\\ \\

1. O.N.Bakharev, M.V.Eremin, and M.A.Teplov, JETP Lett. {\bf 61}, 515 (1995).

2. M.A.Teplov, A.V.Dooglav, E.V.Krjukov et al., JETP {\bf 82}, 370 (1996);
M.A.Teplov, E.V.Krjukov, A.V.Dooglav et al., JETP Lett. {\bf 63}, 227
(1996); O.N.Bakharev, A.V.Dooglav, A.V.Egorov et al., JETP Lett. {\bf 64},
398 (1996); M.A.Teplov, O.N.Bakharev, A.V.Dooglav et al. In: Proc. of the
NATO ASI ''Materials Aspects of High T$_c$ Superconductivity: 10 Years After
the Discovery'' (Delphi, 19-31 Aug. 1996), Kluwer, Dordrecht, in press.

3. Y.Xu and W.Guan, Solid State Comm. {\bf 80}, 105 (1991).

4. A.V.Egorov, Yu.A.Sakhratov, M.A.Teplov et al., to be published.

5. P.W.Anderson, J.Phys.Soc.Jpn. {\bf 9}, 316 (1954); C.P.Slichter,
Principles of Magnetic Resonance, 3-rd ed., Springer-Verlag, New York, 1992;
L.K.Aminov and M.A.Teplov, Sov. Phys. Usp. {\bf 28}, 762 (1985).

\newpage

{\bf Figure captions}\\ \\

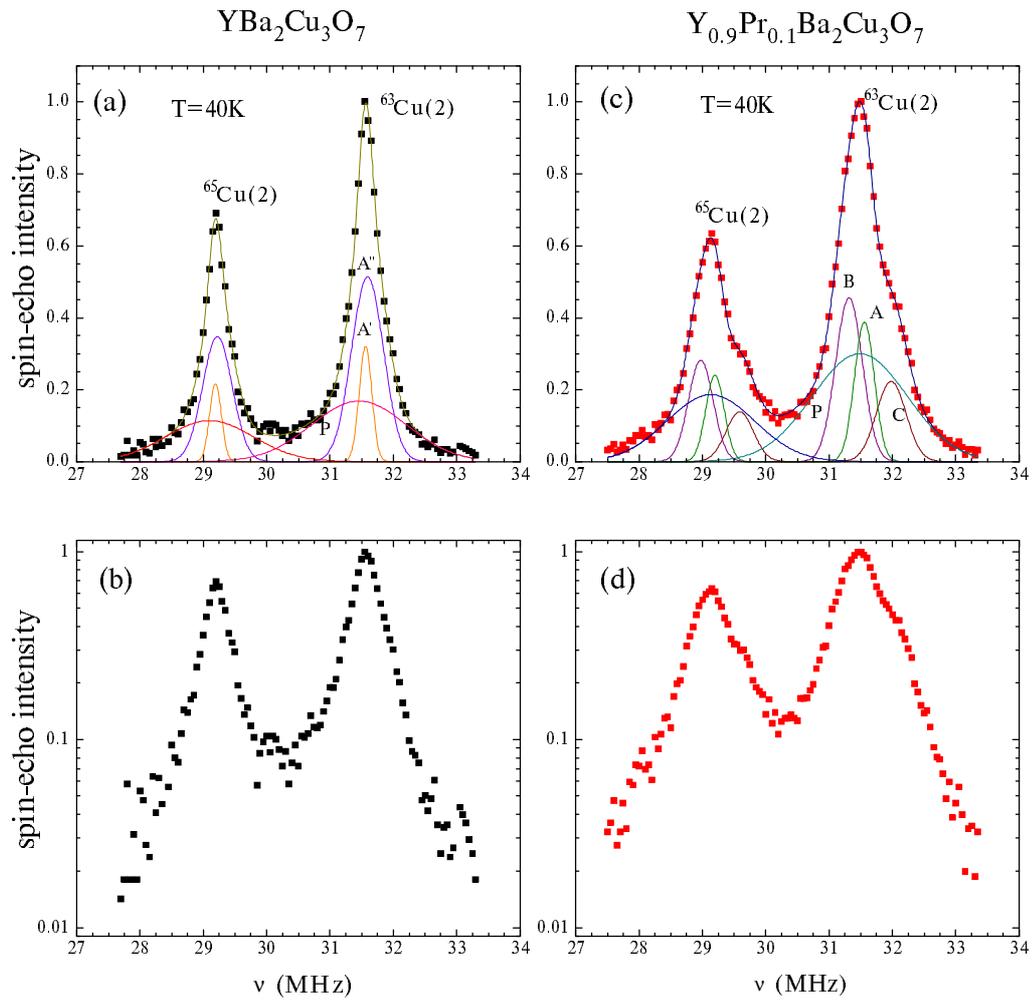
Fig.1. Copper NQR spectra for YBa$_2$Cu$_3$O$_7$ (a,b) and Y$_{0.9}$Pr$%
_{0.1} $Ba$_2$Cu$_3$O$_7$ (c,d) taken at T=40K; solid lines - best fit (see
text and Fig.2).\\

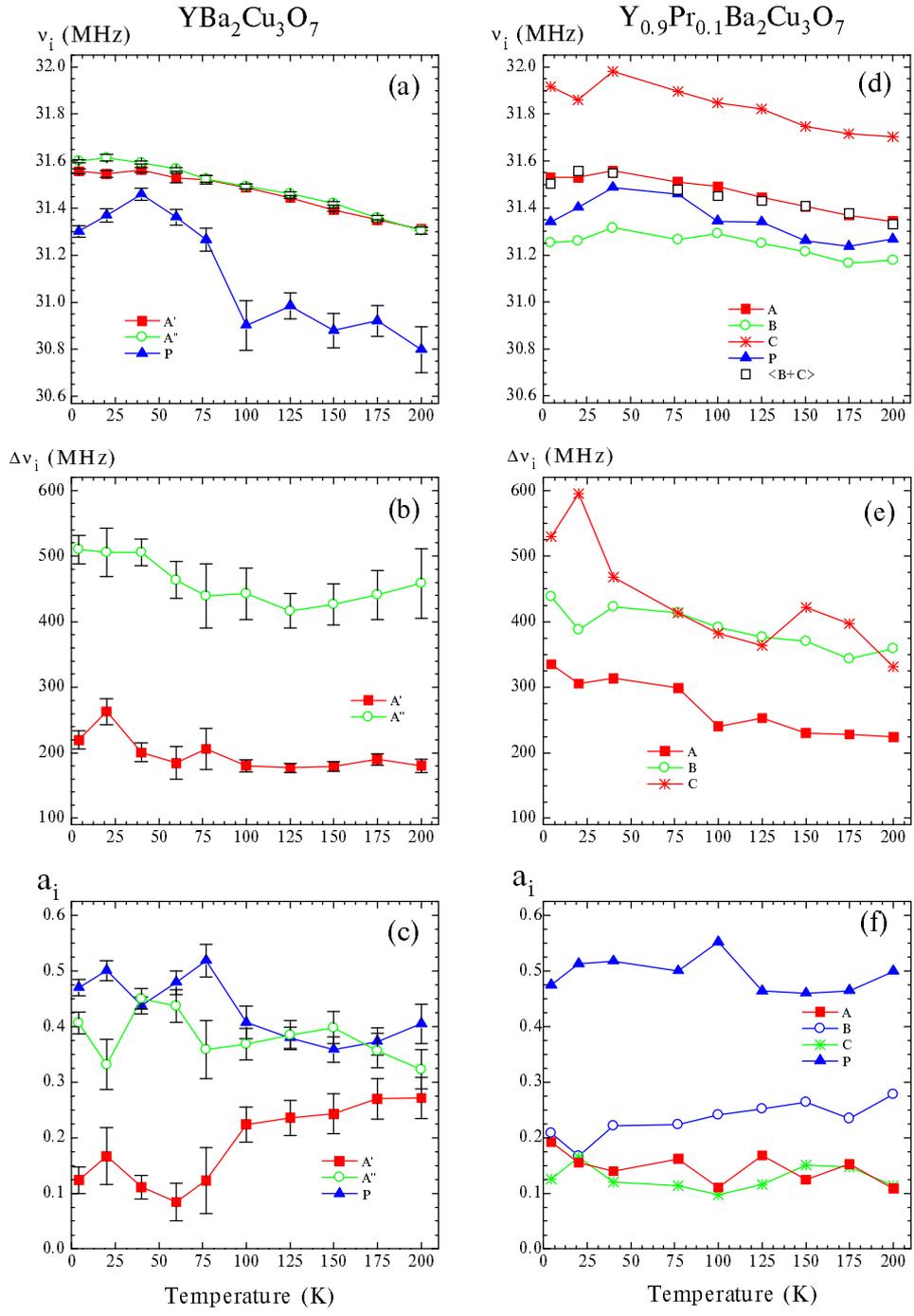
Fig.2. Temperature dependence of parameters of copper NQR spectra for YBa$_2$%
Cu$_3$O$_7$ (a,bc) and Y$_{0.9}$Pr$_{0.1}$Ba$_2$Cu$_3$O$_7$ (d,e,f); for
details see text.\\

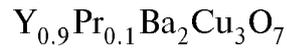
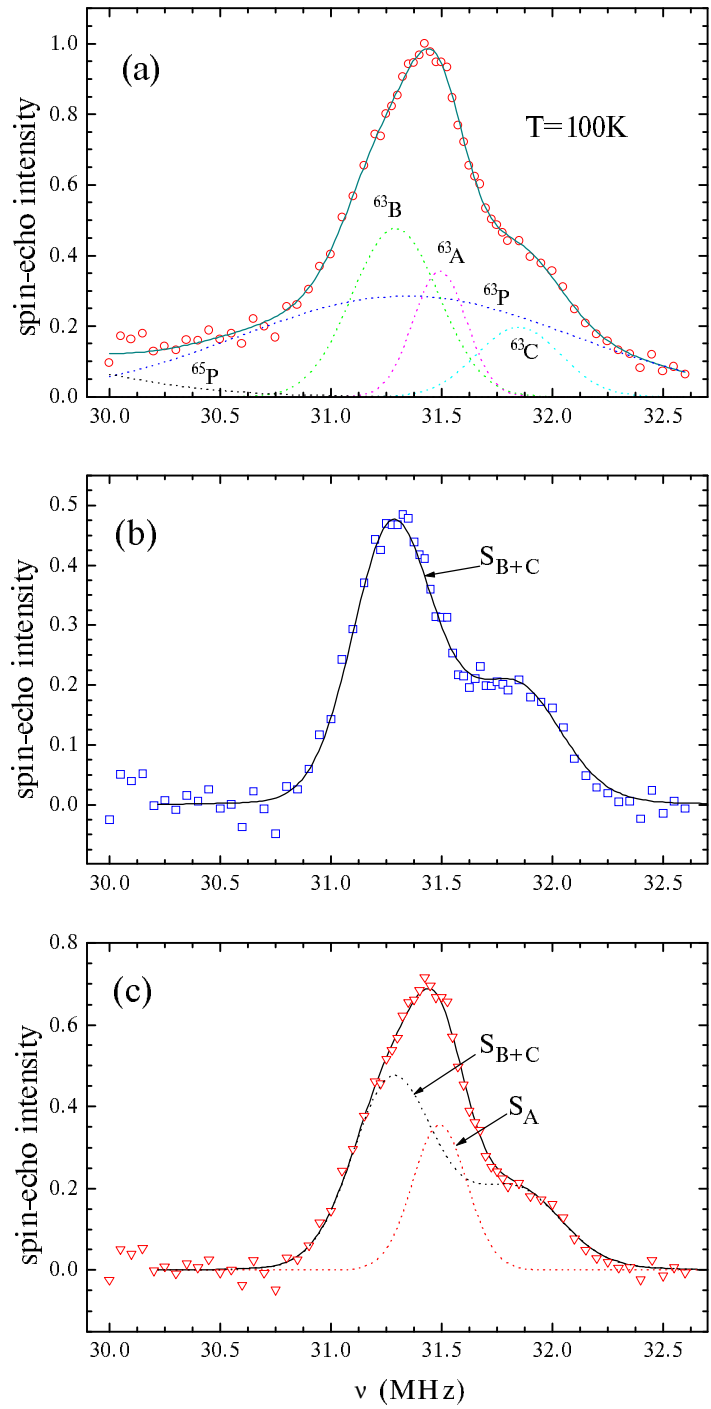
Fig.3. Fragments of $^{63}$Cu(2) NQR spectra for Y$_{0.9}$Pr$_{0.1}$Ba$_2$Cu$%
_3$O$_7$ at T=100K: (a) - best fit to the sum of five Gaussian curves $^{63}$%
($A$+$B$+$C$+$P$) and $^{65}P$; (b) - best fit to the $^{63}$($B$+$C$)
contribution according to Eqs.(1),(2) with parameters $\Delta _B/2\pi =-217$
kHz, $\Delta _C/2\pi =434$ kHz, $\tau _C=2.9$ $\mu $s, $\tau _B=2\tau _C$, $%
\omega _0/2\pi =31.465$ MHz, $\sigma /2\pi =159$ kHz; (c) - the same as (b)
with addition of $^{63}A$ contribution with parameters $\Delta _B/2\pi =-217$
kHz, $\Delta _C/2\pi =434$ kHz, $\tau _C<10^{-7}$ s, $\tau _B=2\tau _C$, $%
\omega _0/2\pi =31.492$ MHz, $\sigma /2\pi =119$ kHz.\\

\end{document}